\documentclass[prd,twocolumn]{revtex4}
\pdfoutput=0
\usepackage{graphicx}
\usepackage{amsmath,amssymb,amsthm,amsbsy}
\usepackage{latexsym}
\usepackage{amsfonts}
\usepackage{amssymb}
\usepackage{tensor}

\usepackage{listings}

\lstdefinestyle{mystyle}{
    basicstyle=\tiny,
    breakatwhitespace=false,         
    breaklines=true,                 
    captionpos=b,                    
    keepspaces=true,                 
    numbers=none,                    
    numbersep=5pt,                  
    showspaces=false,                
    showstringspaces=false,
    showtabs=false,                  
    tabsize=2
}
 
\begin{document}

\title{Quasi-Local Energy of a Charged Rotating Object Described by the Kerr-Newman Metric}
\author{Bjoern S. Schmekel}
\affiliation{Department of Physics, College of Studies for Foreign Diploma Recipients at the University of Hamburg, 20355 Hamburg, Germany}
\email{bss28@cornell.edu}

\begin{abstract}
The Brown-York quasi-local energy of a charged rotating black hole described by the Kerr-Newman metric and enclosed by a fixed-radius surface is computed.
No further assumptions on the angular momentum or the radial coordinate in Boyer-Lindquist coordinates were made. The result can be expressed in terms of 
incomplete elliptic integrals and is used to analyze the self-energy of an electron which is assumed to be described by the Kerr-Newman metric.
For this purpose the small sphere limit is investigated thoroughly comparing the analysis with known results. 
Evaluating the energy using the mass, angular momentum and charge of an electron a value in the order of the Planck energy is obtained in the small sphere limit
as long as the Kerr-Newman metric acting as exterior solution can be used as a description of spacetime. 
\end{abstract}

\maketitle

\section{Introduction}
In this paper the quasi-local energy (QLE) as given by Brown and York \cite{Brown:1992br} of a charged rotating object which is described by the Kerr-Newman metric is calculated.
For this purpose the quasi-local surface energy density is integrated over surfaces with constant $r$ in Boyer-Lindquist coordinates.
The object could be a black hole even though charged black holes play only a minor role in astrophysics. Previous work has dealt with their uncharged counterparts already
\cite{Schmekel:2018wbl} which are more interesting in astrophysical considerations.

The main application is seen in applying the QLE to models which describe elementary particles by the Kerr-Newman metric. Interest in this idea can be traced back at least to
a remark by Carter \cite{PhysRev.174.1559} who pointed out that the magnetic moment of the electron as described by the Dirac equation agrees with the value which can be assigned to this metric (cf. also \cite{PhysRev.174.1559}). 
More recently the idea has been further elaborated on by Burinskii (cf. \cite{Burinskii:2005mm} and the references therein). 
Purely electromagnetic models treating the electron as a charged spinning ring have not proven successful in removing the singular behavior of the self-energy \cite{LyndenBell:2004fk}. 
In classical electrodynamics the reason for the self-energy problem has been well understood (cf. \cite{Low:1997fy,Endres:1992fb}).

Hadrons were also treated as Kerr-Newman black holes \cite{Oldershaw:2007cy}  because of striking similarities \cite{Sivaram1977,Salam1978}.  However, the focus of this article is on leptons only.

The Brown-York QLE has many attractive features which make it stand out among other definitions of energy in general relativity. It can be derived from an action principle
which includes all proper boundary terms, it satisfies a conservation law and possesses a property of additivity. Its value is equal to a Hamiltonian with vanishing shift and unit lapse on the boundary enclosing the region of interest. 
Energy differences within a region can be interpreted  as being caused by a flux of energy into or out of this region. Furthermore, it gives reasonable results in the ADM limit \cite{ADM:1962} - if the metric can be embedded into flat space.
Like energy in classical mechanics it is possible to set a reference energy. In the formalism of the Brown-York QLE this reference point appears as an arbitrary functional
$S_0[\gamma_{ij}]$ in the action. While there is no universal agreement upon how this functional is to be chosen for every possible metric even setting $S_0$ to zero is an 
allowed choice since any choice will leave the underlying equations of motion unchanged. Usually only energy differences are necessary to describe a physical process. 
An absolute reference is only needed when the vacuum itself is subject to being investigated.

The QLE which includes the self-energy of an object has been computed for charged and uncharged non-rotating black holes \cite{Lundgren:2006fu}. In the limit $r \longrightarrow 0$ the negative QLE
approaches the charge $Q$ in the non-rotating case. For uncharged rotating objects the QLE diverges slowly in this limit \cite{Schmekel:2018wbl}. In the latter case the reference term $S_0$ has been omitted. It is conceivable that a
proper reference term would absorb this divergence. In this work a different approach is taken, though, with $S_0$ still being set to zero.

Computer algebra has been used in order to compute the results in this paper with most of the work being done with Maple 18 for Linux and
the add-on package GRTensor III \cite{Pollney:1996kq,GRTensorII}. Some results were double-checked with Mathematica 7 for Solaris 10 and the add-on package MathTensor 2.2.2 \cite{MathTensor}. 

Geometrized units with $G=c=1$ will be used throughout the text unless stated otherwise. 

The paper is organized as follows: First, the framework of the Brown-York quasi-local energy is reviewed briefly. In section III the QLE of the Kerr-Newman spacetime enclosed by a boundary with
fixed radius in Boyer-Lindquist coordinates is computed. In section IV various limits of the Brown-York quasi-local energy are computed connecting to known results. 
Finally, the results are discussed in section V with special emphasis on the small sphere limit which is applied to obtaining the self-energy of the electron. Furthermore, the results
are compared with a small sphere limit and a conservation law which are derived for more general spacetimes. Ultimately though, the result in this limit depends on the details of an interior
solution for the Kerr-Newman spacetime which is an open problem.

\section{Brown-York Quasi-Local energy}

We compute the Brown-York QLE enclosed by a boundary $B$ with induced metric $\sigma_{\mu \nu}$ in a timeslice $\Sigma$ whose time evolution is denoted by $^3 B$ in the form
\begin{eqnarray}
E = \frac{1}{\kappa} \int_B d^2 x \sqrt{\sigma} u_i u_j \tau^{ij} 
\label{BYenergy}
\end{eqnarray}

with the surface stress-energy-momentum tensor being defined as
\begin{eqnarray}
\tau^{ij} & \equiv & \frac{2}{\sqrt{- \gamma}} \frac{\delta S_{cl}}{\delta \gamma_{ij}} = \tau_1^{ij} + \tau_0^{ij} \\ \nonumber
& = & -\frac{1}{\kappa} \left ( \Theta \gamma^{ij} - \Theta^{ij} \right ) - \frac{2}{\sqrt{- \gamma}} \frac{\delta S_0}{\delta \gamma_{ij}}
\label{deftau}
\end{eqnarray}
where $S_{cl}$ is the action consisting of the Einstein-Hilbert term, a potential matter term and boundary terms \cite{PhysRevLett.28.1082}
\begin{eqnarray} \nonumber
S & = & \frac{1}{2 \kappa} \int_M d^4 x \sqrt{-g} \mathcal{R} + \frac{1}{\kappa} \int_{t_i}^{t_f} d^3 x \sqrt{h} K \\ 
& - & \frac{1}{\kappa} \int_{^3 B} d^3 x \sqrt{-\gamma} \Theta - S_0[\gamma_{ij}] + S_m
\label{action}
\end{eqnarray}
evaluated at a classical solution of the Einstein field equations. This effectively suppresses the bulk term and the matter action 
and the definition of $\tau^{ij}$ is based on the presence of the boundary terms. $S_0$ is an arbitrary functional of $\gamma_{ij}$. 
Its inclusion does not alter the equations of motion and is a source of ambiguity. 

The induced metric of $^3 B$ embedded in the spacetime $M$ is labeled $\gamma_{ij}$
and its extrinsic curvature is denoted by $\Theta_{\mu \nu}=-\gamma_{\mu}^{\lambda} \nabla_{\lambda} n_{\nu}$. $\mathcal{D}_\mu t^\nu = \gamma^\alpha_\mu \gamma^\nu_\beta \nabla_\alpha t^\beta$ is
the covariant derivative compatible with $^3 B$. 
The unit normals of $\Sigma$ and $^3 B$ are $u^\mu$ and $n^\mu$, respectively. On $^3 B$ they are assumed to satisfy the orthogonality condition $u \cdot n |_{^3 B} =0$.
$^3 B$ and $B$ share the same normal vector $n^{\mu}$. The surface gravity is denoted by $\kappa$. 
Note that $\tau^{ij}$ includes both the energy due to the gravitational field and the matter fields. 
In general the index "0" refers to reference terms whereas unreferenced quantities are denoted by the index "1".

\section{Evaluation of quasi-local quantities}
We use the Kerr-Newman metric in modified Boyer-Lindquist coordinates
\begin{eqnarray}  
ds^2=- \left ( 1-\frac{2mr-Q^2}{r^2+a^2 \cos^2 \theta} \right ) dt^2 + \nonumber \\
\frac{r^2+a^2 \cos^2 \theta}{r^2-2mr+a^2+Q^2} dr^2 +
\left (r^2 + a^2 \cos^2 \theta \right ) d \theta^2 + \nonumber \\
\sin^2 \theta \left (   r^2 + a^2 + \frac{\left ( 2mr -Q^2 \right ) a^2 \sin^2 \theta}{r^2 + a^2 \cos ^2 \theta}  \right ) d \phi^2 - \nonumber \\
\frac{2a \left ( 2mr -Q^2 \right ) \sin^2 \theta }{r^2 + a^2 \cos^2 \theta} d \phi dt  
\end{eqnarray}

Computing energy and momentum contained in a finite region the results will depend on the chosen boundary. For the remainder of this paper boundaries with $r={\rm const.}$ will be used. The following unit vectors
are chosen

\begin{eqnarray}
u^\mu = \sqrt{\frac{r^2+a^2\cos^2 \theta}{r^2 -2mr +a^2 \cos^2 \theta + Q^2}} \delta_t^\mu \\
n^\mu = \sqrt{\frac{r^2+a^2 -2mr+Q^2}{r^2+a^2 \cos^2 \theta}} \delta_r^\mu
\end{eqnarray}
which satisfy the conditions $n_\mu n^\mu=1$, $u_\mu u^\mu=-1$ and $u_\mu n^\mu=0$. Furthermore, $u_{\mu} \gamma^{\mu \nu}=u^{\nu}$ and $n_{\mu} \gamma^{\mu \nu}=0$
since $\gamma_{\mu \nu} = g_{\mu \nu} - n_{\mu} n_{\nu}$. 

Evaluating with the aid of computer algebra yields

\begin{widetext}

\begin{eqnarray}
\det{\sigma} = -{\frac { \left( {Q}^{2}+{a}^{2}-2\,mr+{r}^{2} \right)  \left( {\chi}^
{4}{a}^{4}+2\,{\chi}^{2}{a}^{2}{r}^{2}+{r}^{4} \right)  \left( {\chi}^
{2}-1 \right) }{{a}^{2}{\chi}^{2}+{Q}^{2}-2\,mr+{r}^{2}}}
\end{eqnarray}

\begin{eqnarray} \nonumber
\epsilon  \equiv  u_i u_j \tau^{ij}=
 - \frac{1}{{\left( {{Q^2} + {a^2} - 2\,mr + {r^2}} \right)\kappa \,\left( {{a^2}{\chi ^2} + {Q^2} - 2\,mr + {r^2}} \right)\left( {{a^2}{\chi ^2} + {r^2}} \right)}}\sqrt {\frac{{{Q^2} + {a^2} - 2\,mr + {r^2}}}{{{a^2}{\chi ^2} + {r^2}}}}  \cdot \\ \nonumber
\left[ { - {\chi ^4}{a^4}m + {\chi ^4}{a^4}r + 2\,{\chi ^2}{Q^2}{a^2}r + {\chi ^2}{a^4}m + {\chi ^2}{a^4}r - 5\,{\chi ^2}{a^2}m{r^2} + 3\,{\chi ^2}{a^2}{r^3}} \right . \\
\left . {+ 2\,{Q^4}r + 2\,{Q^2}{a^2}r - 8\,{Q^2}m{r^2} + 4\,{Q^2}{r^3} - 3\,{a^2}m{r^2} + {a^2}{r^3} + 8\,{m^2}{r^3} - 8\,m{r^4} + 2\,{r^5}} \right]
\label{epskerrnewmanmaple}
\end{eqnarray}

\begin{eqnarray} \nonumber
j_{\phi} \equiv -\delta_{\phi}^a \sigma_{ai} u_j \tau^{ij} =  {\frac {a \left( {\chi}^{2}{a}^{2}m+{Q}^{2}r-m{r}^{2} \right)  \left( 
{\chi}^{2}-1 \right) \sqrt {{Q}^{2}+{a}^{2}-2\,mr+{r}^{2}}}{ \left( {a
}^{2}{\chi}^{2}+{Q}^{2}-2\,mr+{r}^{2} \right) ^{3/2} \left( {a}^{2}{
\chi}^{2}+{r}^{2} \right) \kappa}} \\ 
\end{eqnarray}

\end{widetext}
where the result for the single non-vanishing component of the quasi-local momentum surface density $j_{\phi}$ has been added for the sake of completeness only and is not needed in subsequent steps. 
Intermediate results for $\Theta^\mu_\nu$ and $\tau^{\mu \nu}$ can be found in the appendix.
Using the variable substitution $\chi=\cos \theta$ the integration over $d \theta$ giving the QLE
\begin{eqnarray}
E=2 \pi \int_0^{\pi} d \theta \sqrt{\sigma} \epsilon = 2 \pi \int_{-1}^{1} d \chi \frac{d \theta}{d \chi} \sqrt{\sigma} \epsilon
\label{intE}
\end{eqnarray}
succeeds using Maple. This results in a complex expression which can be expressed in terms of the incomplete elliptic integrals
\begin{eqnarray}
\mathfrak{E}(z,k) \equiv \int_0^z \frac{\sqrt{1-k^2 \zeta^2}}{\sqrt{1-\zeta^2}} d \zeta \\
\mathfrak{F}(z,k) \equiv \int_0^z \frac{1}{\sqrt{1-\zeta^2}\sqrt{1-k^2 \zeta^2}} d \zeta
\end{eqnarray}
With
\begin{eqnarray}
\tilde \Xi_E \equiv \mathfrak{E} \left( i \left | \frac{a}{r} \right | , \frac{|r|}{\sqrt{Q^2-2mr+r^2}} \right ) \\
\tilde \Xi_F \equiv \mathfrak{F} \left( i \left | \frac{a}{r} \right | , \frac{|r|}{\sqrt{Q^2-2mr+r^2}} \right )
\end{eqnarray}
we obtain
\begin{eqnarray}
-i\frac{{6m{r^2} - 2{r^3} - r\left( {4{m^2} + 2{Q^2} + {a^2}} \right) + 2m{Q^2} + m{a^2}}}{{2\left| a \right|\sqrt {{Q^2} - 2mr + {r^2}} }}{\tilde \Xi _E} + \nonumber \\
i\frac{{m{r^2} + {r^3} - r\left( {4{m^2} - {a^2}} \right) + 2m{Q^2} + m{a^2}}}{{2\left| a \right|\sqrt {{Q^2} - 2mr + {r^2}} }}{\tilde \Xi _F} - \nonumber \\
\frac{{\left( {m - r} \right)\sqrt {\left( {{r^2} + {a^2}} \right)\left( {{Q^2} + {a^2} - 2mr + {r^2}} \right)} }}{{2{Q^2} - 4mr + 2{r^2}}} = E{}_1 \nonumber \\
 \label{Enoref}
\end{eqnarray}

unless $\sqrt{r(2m-r)-Q^2}<|a|$ and $r(2m-r)-Q^2 \ge 0$. If this condition is met the integral diverges. Eqn. \ref{Enoref} may be used to analytically continue the QLE into this undefined region
if analyticity can be imposed on the QLE. The resulting expression can be evaluated numerically. Suitable Maple input code for $E_1$ given by eqn. \ref{Enoref} can be found in the appendix. 

\section{Limits of $E_1$}
\subsection{ADM limit $r \longrightarrow \infty$}
Employing the relations
\begin{eqnarray}
\mathfrak{E}(x,1) & = & x 
\\
\mathfrak{F}(x,1) & \approx & x + \frac{1}{3} x^3
\end{eqnarray}
with the latter being valid for small values of $x$ eqn. \ref{Enoref} can be expressed as an asymptotic expansion in the limit  $r \longrightarrow \infty$. 
We obtain
\begin{eqnarray}
E_1 = m -r + O(r^{-1})
\end{eqnarray}
This would exactly give the expected ADM mass \cite{ADM:1962} $E \longrightarrow m$ if the subtraction term $E_0=-r$ was to be subtracted. 

\subsection{$a=0$}
Using the expansions
\begin{eqnarray}
\mathfrak{E}(x,z) & \approx & x 
\\
\mathfrak{F}(x,z) & \approx & x 
\end{eqnarray}
for small values of $x$ eqn. \ref{Enoref} gives  for $r>0$ \cite{Lundgren:2006fu}
\begin{eqnarray}
E_1 = -r \sqrt{1-\frac{2m}{r}+\frac{Q^2}{r^2}}
\label{EnorefLS}
\end{eqnarray}
which further reduces to $E_1 \approx -Q$ in the limit $r \longrightarrow 0$ or
\begin{eqnarray}
E_1 = -r \sqrt{1-\frac{2m}{r}}
\end{eqnarray}
in the Schwarzschild limit $a=0$, $Q=0$ \cite{Brown:1992br} with $\kappa = 8 \pi$. 

\subsection{$Q=0$}

For $Q=0$ eqn. \ref{Enoref} reduces to the case without charge \cite{Schmekel:2018wbl}
\begin{eqnarray}
 + \frac{{i\left| r \right|\left[ {\left( {6m - 2r} \right){{\left| r \right|}^2} - \left( {4{m^2} + {a^2}} \right)r + {a^2}m} \right]}}{{4\left| a \right|r\left( {m - \frac{r}{2}} \right)}}{\Xi _E} \nonumber \\
 + \frac{{i\left| r \right|\left[ {\left( {5m - r} \right){{\left| r \right|}^2} - \left( {6{m^2} + {a^2}} \right)r + 3{a^2}m} \right]}}{{4\left| a \right|r\left( {m - \frac{r}{2}} \right)}}{\Xi _F} \nonumber \\
 - \frac{{\left( {m - r} \right)\sqrt {{a^2} + {r^2}} \left[ {r\left( {2m - r} \right) - {a^2}} \right]}}{{4r\left( {m - \frac{r}{2}} \right)\sqrt {{a^2} - 2mr + {r^2}} }} = E_1
 \label{EnorefQ0}
\end{eqnarray}
where
\begin{eqnarray}
\Xi_E \equiv \mathfrak{E} \left( \left| a \right| \sqrt {{\frac {1}{r\, \left( 2\,m-r \right) }}} ,\sqrt {1-\frac{2m}{r}} \right) \\
\Xi_F \equiv \mathfrak{F} \left( \left| a \right| \sqrt {{\frac {1}{r\, \left( 2\,m-r \right) }}}  ,\sqrt {1-\frac{2m}{r}} \right)
\end{eqnarray}
This expression can be obtained in the limit $Q=0$ from eqn. \ref{Enoref} using the relations $\mathfrak{F}(z,k) = k^{-1} \mathfrak{F}(zk,k^{-1})$ and
\begin{eqnarray}
\mathfrak{E} \left ( z,k  \right )  = k \mathfrak{E} \left ( zk,k^{-1}   \right ) + \frac{1-k^2}{k} \mathfrak{F} \left ( zk,k^{-1}  \right ) 
\end{eqnarray}
Combining the last relation with the former multiplied by another factor $x$ we obtain the desired result
\begin{eqnarray} \nonumber
\mathfrak{E} \left ( \frac{p}{k},k  \right ) + x \mathfrak{F} \left ( \frac{p}{k},k  \right ) = k \mathfrak{E} \left ( p,\frac{1}{k}  \right ) + \frac{1+x-k^2}{k} \mathfrak{F} \left ( p,\frac{1}{k}  \right )  \\
\end{eqnarray}
which can be used to convert eqn. \ref{Enoref} into eqn. \ref{EnorefQ0}. In astrophysical applications where the energy of an uncharged rotating black hole outside its event horizon
is considered we prefer the functions $\Xi_E$ and $\Xi_F$ used in eqn. \ref{EnorefQ0} because the arguments of the elliptic integrals become real. 

\subsection{$r \longrightarrow 0$ with $m \ll Q \ll a$}
In the small sphere limit additionally imposing $Q \ll a$ eqn. \ref{Enoref} simplifies to
\begin{eqnarray}
E_1 
\approx - \frac{m}{2} \left [ i \left | \frac{a}{Q} \right | \left ( \tilde \Xi_E - \tilde \Xi_F \right ) + \left | \frac{a}{Q} \right | ^2 \right ]
\label{E1r0limit}
\end{eqnarray}

The difference of the two incomplete elliptic integrals can be written as
\begin{eqnarray}
\nonumber
\tilde \Xi_F - \tilde \Xi_E & = & \frac{r^2}{Q^2} \int_0^{\frac{ia}{r}} d \zeta \frac{\zeta^2}{\sqrt{1-\zeta^2} \sqrt{1-\frac{r^2}{Q^2} \zeta^2}} \\ \nonumber
& = & \frac{r^2}{Q^2} \cdot \left ( \frac{ia}{r} \right )^3 \int_0^1 d \zeta^{\prime} \frac{\zeta^{\prime 2}}{\sqrt{1+\frac{a^2}{r^2}  \zeta^{\prime 2}} \sqrt{1 + \frac{a^2}{Q^2} \zeta^{\prime 2}} } \\
\end{eqnarray}
using their definitions. Expanding the integrand about $r=0$ and performing the integration gives
\begin{eqnarray} \nonumber
\tilde \Xi_F - \tilde \Xi_E = -i \frac{Q^2 + a^2 - Q \sqrt{Q^2+a^2}}{Q \sqrt{Q^2 + a^2}} \approx i \left ( 1 - \left | \frac{a}{Q} \right | \right )
\\
\label{approxXiFXiE}
\end{eqnarray}
Finally,
\begin{eqnarray}
E_1 \approx - \frac{m}{2} \left |  \frac{a}{Q} \right | 
\label{Enorefr0}
\end{eqnarray}
Note that despite the ratio $a/Q$ being huge for the application depicted in fig. \ref{E_self_electron} the contribution from the first term in the parenthesis of eqn. \ref{approxXiFXiE} cannot be neglected because
the second terms cancels the $|a/Q|^2$ term in eqn. \ref{E1r0limit}. This makes finding a good approximation for the integrals in $E_1$ in this limit difficult because the desired $|a/Q|$ term is overshadowed by 
a contribution from a $|a/Q|^2$ term. Numerical evaluation of eqn. \ref{Enoref} (cf. fig. \ref{E_self_electron}) is challenging requiring arbitrary precision arithmetics.

\section{Results and Discussion}
\subsection{Description of Elementary Particles by the Kerr-Newman Metric}
As stated before the final result given in eqn. \ref{Enoref} should be employed to compute the self-energy of an electron with a brief outlook to other fermions. 
One may argue how reasonable it is to treat an electron as a "black hole". The term "black hole" should not be taken too literally, though. Due to the 
values of $a$, $Q$ and m of the electron any such blackhole would be super-extremal exposing a naked ring singularity which is not hidden behind an event horizon. 
While the presence of naked singularities may be disturbing there appears to be no fundamental argument to rule them out. In fact, numerical solutions 
of the Einstein field equations indicate that the formation of naked singularities starting from reasonable initial data is possible \cite{PhysRevLett.66.994}. A proper interior solution matched against the Kerr-Newman
metric may lead to a spacetime free of singularities, though. 

Another problem is the Compton wavelength of the electron which is 
many orders of magnitude larger than the classical Schwarzschild radius which raises doubts as to whether the electron "would fit inside" the horizon if there was a horizon to begin with.  
However, up to now the correct theory of quantum gravity is still unknown, so whether this argument can be upheld can presently not be decided. The work in this paper is purely classical
in nature with $\hbar$ (in a suitable unit system) entering the model via the angular momentum of the electron only. 

In QED interactions are assumed to be point-like which is the cause for divergent self-energy terms. As it is well known the problem is not solved by the fuzziness of quantum 
mechanics but rather by a proper renormalization procedure accompanied by QED with the latter providing structure to the electron surrounding it by a cloud of virtual electron-positron pairs. 
The Kerr-Newman metric might provide the electron with the necessary structure given that the usual renormalization procedures do not work for the gravitational field. 

Computing the QLE a reference energy has to be chosen. As in past treatments this reference term will be omitted setting $S_0$ to zero which
is harder to justify in this context because we will possibly be making statements about the vacuum. Since most statements will rely on results in the small sphere limit
$r \longrightarrow 0$ we assume that a proper reference term, e.g. obtained by some sort of counter-term method \cite{Astefanesei:2006zd,Clarkson:2002uj}, will vanish in this limit. Furthermore, due to the relation \cite{Schmekel:2018wbl}
\begin{eqnarray} \nonumber
\int_{t_i \cap ^3 B}^{t_f \cap ^3 B} d^2 x \sqrt{\sigma} u_i u_j \tau^{ij}   = 
& - & \int_{^3 B} d^3 x \sqrt{-\gamma} \tau^{ij} \mathcal{D}_i u_j 
\\ \nonumber
& + & \int_{^3 B} d^3 x \sqrt{-\gamma} u_\mu n_\nu T^{\mu \nu}
\\
\label{conslaw}
\end{eqnarray}
which acts as a statement of conservation of energy
we could consider a hypothetical process in which an existing Schwarzschild black hole is fed through a flux of stress energy until it reaches the desired values of mass, angular momentum and charge. 
The initial Schwarzschild black hole serves as a reference whose QLE (unreferenced or referenced by an embedding into flat space) approaches zero as $r \longrightarrow 0$. 
For the Kerr-Newman metric and our choices of normal vectors $\tau_1^{ij} \mathcal{D}_i u_j$ vanishes. 

In fig. \ref{E_self_electron} the QLE is plotted as a function of $r$ for the mass, angular momentum and charge of an electron. In the limit $r \longrightarrow 0$ the QLE converges to
\begin{eqnarray}
\mathop {\lim }\limits_{r \to 0} {E_1} =  - 2.926 \cdot {E_{{\rm{Planck}}}}
\label{E_self}
\end{eqnarray}
 The same result is obtained from eqn. \ref{Enorefr0} which can be further approximated in the limit $Q \ll a$
 \begin{eqnarray}
E_1 \approx - \frac{m}{2}  \frac{| a |}{| Q |} 
\end{eqnarray}
Note that in order to satisfy the last condition $m \ll J / Q$ has to be satisfied as well since $a=J/m$. 
Also, eqn. \ref{EnorefQ0} predicts $E_1$ to diverge for $a \neq 0$ and $Q=0$. Thus, in this model a particle which has spin needs to possess charge as well.
If we insist on the absence of a horizon we also have to demand $a^2+Q^2>m^2$. 
No significant deviations from the characteristics of the plot in fig. \ref{E_self_electron} can be seen for the mass of the muon or tau with the only difference being the saturation energy given
by eqn. \ref{Enorefr0} which becomes visible below $r < 10^{-55} {\rm cm} $ and  $r < 10^{-53} {\rm cm} $, respectively. 

The picture for the neutrino is more involved. For $m=0$ and $r \longrightarrow 0$ the conditions for eqn. \ref{Enoref} to be valid are satisfied, but for $m \neq 0$ and $r \longrightarrow 0$ they are not. 
Despite the convergence conditions not being satisfied evaluating eqn. \ref{Enoref} (thus analytically continuing the expression into the invalid region) results in huge but finite values for $E_1$. 
 More work is needed to fully understand this case - possibly considering a non-zero cosmological constant as well.  
 
 Fig \ref{E_self_electron} suggests that the values of the mass, charge and angular momentum
 are such that the resulting self-energy is neither much below nor much beyond the Planck energy. Thus, we speculate that the electron can be considered to be the fermion   
 whose self-energy at $r=0$ is in the order of the Planck energy. On sub-atomic scales the Planck energy is huge. In this context it is interesting to note that the Planck energy can be computed
 as the mass which produces a black hole whose Compton wavelength is equal to its Schwarzschild radius, thus providing a possible way out of the problem of the Compton length of the electron
 being much larger than its Schwarzschild radius (cf. also \cite{Carr:2015nqa,Ha:2009kg,DennyLee:2015} ). 
 
 For large values of $r$ the unreferenced QLE $E_1$ approaches $m-r$. If one could use the reference term $E_0 = -r$ for the Schwarzschild case one would obtain the ADM mass \cite{ADM:1962} in this limit.

 The muon and the tau are unstable particles and could possibly be described as excited states of
 a yet unknown theory. It is pointed out again that the presented model is purely classical in nature in the sense that it is not quantized in any way. $\hbar$ enters the model as 
 parameter of the angular momentum only. 
 So far no distinction between the measured renormalized values for charge and mass and their corresponding bare values has been made using the former 
 throughout this work without exception. It may be argued that for the proposed application the bare values would have to be used since the model is supposed
 to describe a single particle. This issue may be resolved by the fact that the QLE "measures" the energy within a region enclosed by a boundary which would
 possibly include clouds of virtual particles. Due to the considered solutions being super extremal the particles would not be shielded by an event horizon. 
 Whether such superpositions are feasible has to be investigated in a future work. Because of the non-linear nature of general relativity the results are not obvious. 
 
\subsection{Small Sphere Limit}
This maybe somewhat unexpected result presented in the previous paragraph may bring up questions as to how reliable the Brown-York QLE is on small length scales and whether
the concept of QLE may break down altogether. While there is still no universal consensus on how (quasi-local) energy is to be defined in general relativity it shall be emphasized again
that the QLE used in this work follows from careful Hamilton-Jacobi analysis with proper treatment of all necessary boundary terms in the action. Therefore, the Brown-York QLE will break down
when general relativity itself breaks down. When and whether this is the case is an open question. Nevertheless, the Brown-York QLE is known to give reasonable results in the small
sphere limit \cite{Brown:1998bt}. Because of the absence of a local energy-momentum contribution from the gravitational field the QLE due to the latter vanishes with the only non-vanishing contribution
to the QLE contained in a small sphere with a radius close to zero coming from the stress-energy content. Indeed, the results derived in the small sphere limit \cite{Brown:1998bt} indicate that in the presence of matter
\begin{eqnarray}
E_1=-\hat r + \frac{1}{18} \hat r^3 \left [ 5 R_{\mu \nu}  u^{\mu} u^{\nu} + 2R \right ]_P + O(\hat r^4)
\label{SmallSphereMatterE1}
\end{eqnarray}
\begin{eqnarray}
E=\frac{4}{3} \pi \hat r^3 \left [ T_{\mu \nu} u^{\mu} u^{\nu} \right ]_P + O(\hat r^4)
\label{Er0matter}
\end{eqnarray}
and in the absence of matter to lowest order
\begin{eqnarray}
E_1 = -\hat r + \frac{7}{450} \hat r^5 \left [ T_{\mu \nu \rho \sigma} u^{\mu} u^{\nu} u^{\rho} u^{\sigma} \right ]_P + O(\hat r^6)
\label{SmallSphereVacuumE1}
\\
E= \frac{1}{90} \hat r^5 \left [ T_{\mu \nu \rho \sigma} u^{\mu} u^{\nu} u^{\rho} u^{\sigma} \right ]_P + O(\hat r^6)
\end{eqnarray}
where $\hat r$ is an affine parameter along the generators of the lightcone belonging to the point $P$ where $[...]_P$ denotes evaluation at $P$. In these equations the Bel-Robinson tensor \cite{Bel:1958,Bel:1959,Robinson:1958,Garecki:2000dj} being defined as
\begin{eqnarray}
T \indices{_\mu _\nu _\rho _\sigma} \equiv C \indices{_\mu _\lambda _\rho _\kappa} C \indices{_\nu  ^\lambda  _\sigma  ^\kappa} + {^*}C \indices{_\mu _\lambda _\rho _\kappa} {^*}C \indices{_\nu ^\lambda _\sigma ^\kappa }
\end{eqnarray}
is used where $C_{\mu \nu \rho \sigma}$ is the Weyl tensor and $^*C \indices{_\mu _\nu _\rho _\sigma} = \epsilon \indices{_\mu _\nu _\alpha _\beta} C \indices{^\alpha ^\beta  _ \rho _\sigma}$ is its left dual. 

According to eqn. \ref{Er0matter} the QLE vanishes in the limit $r \longrightarrow 0$ unless $T_{\mu \nu}$ is singular at $r=0$ in which case the result is harder to predict.

\begin{widetext}

\begin{figure}
\scalebox{0.8}{\includegraphics{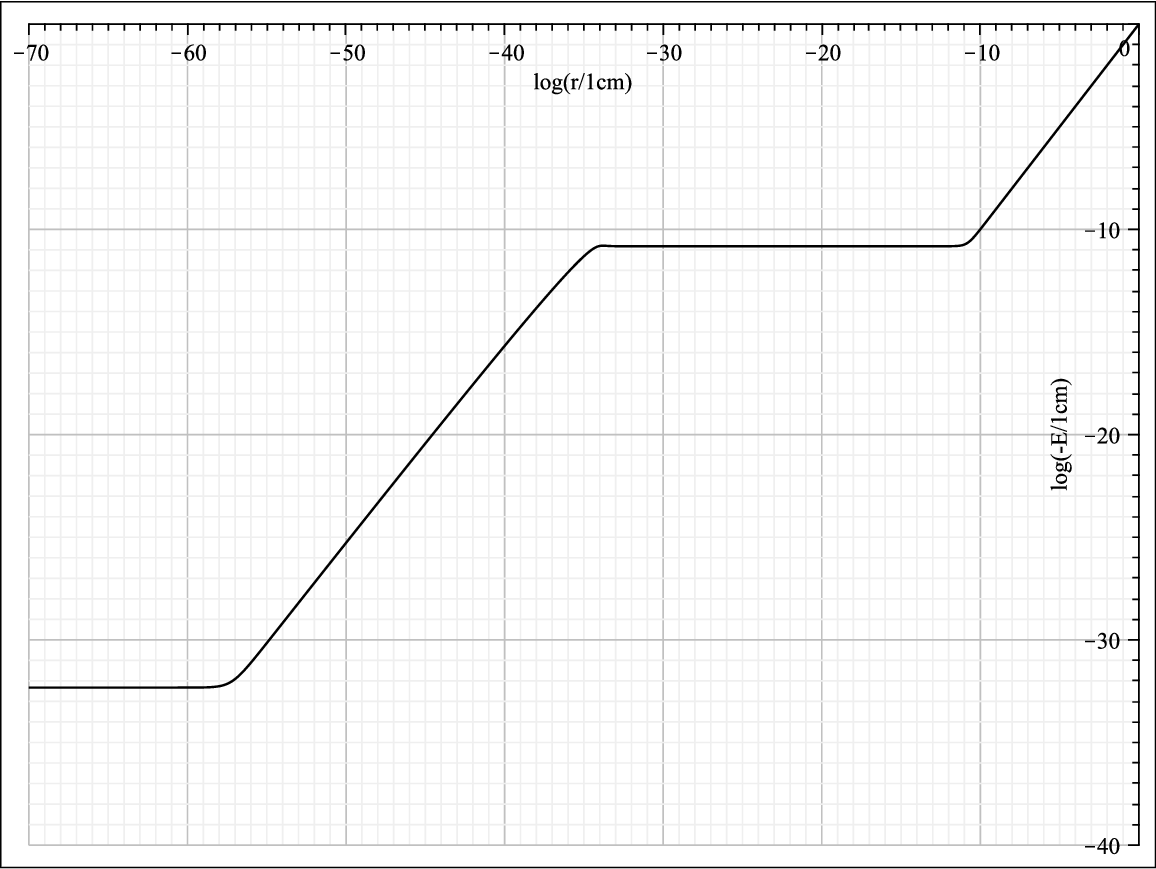}}
\caption{$\log (-E / 1{\rm cm})$ vs. $\log (r / 1{\rm cm})$:
Note the change of behavior at the different length scales $r_p=1.6 \cdot 10^{-33} {\rm cm}$ (Planck length) which is in rough agreement with the scale associated with the charge
$r_q=Q=1.38 \cdot 10^{-34} {\rm cm}$ and $r_a=a=J/m=1.93 \cdot 10^{-11} {\rm cm}$ which is roughly in the order of the Compton wavelength  $r_c=2.426 \cdot 10^{-10} {\rm cm}$
and also provides the size of the ring singularity present in the Kerr-Newman metric. Furthermore, the classical Schwarzschild radius is given by $r_s = 2m = 1.35  \cdot 10^{-55} {\rm cm}$. 
For large values of $r$ the unreferenced QLE approaches $m-r$. For small values of $r$ the negative QLE approaches approximately three times the Planck energy eqn. \ref{E_self}.  Also,
it is interesting to note that $r_c / r_a = 4 \pi \approx 12.57$ and $r_p / r_q \approx 11.59$.  
}
\label{E_self_electron}
\end{figure}

\end{widetext}

The unreferenced QLE $E_1$ depicted in fig. \ref{E_self_electron} approaches the constant value given in eqn. \ref{E_self}. Since $E_1$ remains constant below $r<10^{-58} {\rm cm}$ the contribution
from the gravitational field energy has dropped out below this length-scale. Thus, the huge value must be due to a contribution from stress-energy alone. Note, that $E_1(0)\neq 0 $ is not a contradiction. 
With the Kerr(-Newman) metric belonging to the class of (electro-)vacuum solutions $T_{\mu \nu}$ is assumed to be zero in the absence of charge everywhere except at $r=0$. For a metric describing the full spacetime
the Kerr-Newman solution would have to be matched against an interior solution at a yet unknown position $r=r_i$. How such an interior solution for the Kerr-Newman spacetime could be obtained remains an open problem. 
Despite this uncertainty down to $r=r_i$ the details of the interior solution have no impact on the QLE in the region $r \ge r_i$ with the QLE being defined as a surface integral over a surface stress-energy momentum tensor. 
This includes the possibility of the Kerr-Newman metric being the correct description for the entire spacetime with the exception of an open ball around $r=0$. Within this open ball the QLE would drop off to zero as 
$r \longrightarrow 0$ in accordance with eqn. \ref{Er0matter}. 

As mentioned before a small pit of stress-energy in the order of the energy given in eqn. \ref{E_self} would have the interesting property of possessing a Compton wavelength in the order of the Schwarzschild radius associated with this energy,
so the pit would fit inside its own Compton length.

Alternatively, the local absence of a contribution from the gravitational field to the QLE can be inferred by inspection of eqn. \ref{conslaw}. Consider a region of empty space bounded by the boundary $B$, i.e. a region which does not contain any stress-energy or momentum and which is very far away from other sources such that contributions from gravitational fields can be neglected. The QLE in that region will just increase by the amount of stress-energy which flows into this region if the first term on the right hand side of eqn. \ref{conslaw} is zero. 

Expanding the covariant derivatives in Riemann normal coordinates as
\begin{eqnarray}
\nabla_k n_l = \partial_k n_l - \frac{1}{3} \left [ R_{lkmp} + R_{lmkp} \right ]_P n^m x^p
\end{eqnarray}
and similarly for $\nabla_k u_l$ at a small distance $x^p = r \cdot n^p$ from the point $P$ and dropping partial derivatives in subsequent steps for our choice of normal vectors the first term on the right hand side 
of eqn. \ref{conslaw} can be written as
\begin{eqnarray} \nonumber
& - & \int_{^3 B} d^3 x \sqrt{-\gamma} \tau^{ij} \mathcal{D}_i u_j = \\ \nonumber
& - &  \frac{1}{\kappa} \int_{^3 B} d^3 x \sqrt{-\gamma} \left ( \mathcal{D}_i u_j \right ) \left ( \gamma^{ij} \gamma^{kl} - \gamma^{ik} g^{jl} \right ) \nabla_k n_l = \\ \nonumber
& - & \left ( - \frac{r}{3} \right )^2 \frac{1}{\kappa}  \int_{^3 B} d^3 x \sqrt{-\gamma}  
\left [ R \indices{^i ^l} R \indices{^b ^j}  - R \indices{_a ^i _c ^l } R \indices{^a ^b ^c ^j} \right ]_P n_{i} n_{l} n_b u_j
\\
\label{conslawR}
\end{eqnarray}
Note that this relation is a geometrical result which is valid even without invoking the Einstein field equations. For some of the required manipulations
the computer algebra system Cadabra has been proven useful \cite{Peeters:2007wn,Peeters2018Cadabra2}. 

In the absence of matter characterized by $T_{\mu \nu} = 0$ the Einstein field equations allow us to set $R_{\mu \nu}=0$ and to replace all occurrences of the Riemann tensor with the Weyl tensor.
Using the Bel-Robinson tensor eqn. \ref{conslawR} can be written as
\begin{eqnarray}
& - & \int_{^3 B} d^3 x \sqrt{-\gamma} \tau^{ij} \mathcal{D}_i u_j = \\ \nonumber
& + &   \frac{r^2}{18 \kappa}   \int_{^3 B} d^3 x \sqrt{-\gamma}  
\left [ n_{i} n_{l} n_b u_j T^{l j i b} \right ]_P 
\label{conslawC}
\end{eqnarray}
where the Bel-Robinson tensor has been expressed as
\begin{eqnarray}
T \indices{^l ^j ^i ^b} = C \indices{^l ^c ^a ^i} C \indices{^j _c _a ^b} + C \indices{^l ^c ^a ^b} C \indices{^j _c _a ^i} - \frac{1}{2} g \indices{^l ^j} C \indices{^c ^a ^f ^i} C \indices{_c _a _f ^b}
\end{eqnarray}
using the standard symmetries of the Weyl tensor. 

Eqn. \ref{conslaw} together with eqn. \ref{conslawR} or eqn. \ref{conslawC} provides a result which is similar to the small sphere limit by Brown, Lau and York expressed by eqns. \ref{SmallSphereMatterE1} and \ref{SmallSphereVacuumE1}.  Whereas the latter relate the QLE  to the stress-energy contained in a small region of interest or a contribution of field energy expressed by the Bel-Robinson tensor in the absence of matter eqn. \ref{conslaw} relates the difference of QLE at two different values of $t$ to a flux of stress-energy into or out of a small region of interest and a contribution of field energy expressed by the Bel-Robinson tensor in the absence of matter. Since in both eqn. \ref{conslawR} and eqn. \ref{conslawC} the contribution is expressed as a surface integral of a quantity which is a contraction with $n_{\mu}$ over the boundary $^3 B$ this contribution can be interpreted as a flux into or out of $B$ which is not immediately apparent in eqn. \ref{conslaw}. Like in classical physics the statement of conservation of energy is expressed by an equality of a time derivative and a flux of energy through a closed surface. In the present case the flux consists of two terms with one of them describing the flux of matter. The second term is due to a flux of gravitational energy (cf. eqn. \ref{conslawC}) and another contribution due to
a flow of stress-energy provided by the product of the Ricci tensors in eqn. \ref{conslawR} which are higher ordered in $r$, though. This observation is another reason why the QLE defined by Brown and York ought to be regarded
as the total energy contained within $B$. 
Even when the region of interest is not small the change in QLE can still be written as a flux. Starting from the second line in eqn. \ref{conslawR} we obtain using the product rule
\begin{eqnarray} \nonumber
& - & \int_{^3 B} d^3 x \sqrt{-\gamma} \tau^{ij} \mathcal{D}_i u_j = \\ \nonumber
& + &  \frac{1}{\kappa} \int_{^3 B} d^3 x \sqrt{-\gamma}  \nabla_k  \left [ \left ( \mathcal{D}_i u_j \right ) \left ( \gamma^{ij} \gamma^{kl} - \gamma^{ik} g^{jl} \right ) \right ] n_l  \\ \nonumber
& - &  \frac{1}{\kappa} \int_{^3 B} d^3 x \sqrt{-\gamma}  \nabla_k  \left [ \left ( \mathcal{D}_i u_j \right ) \left ( \gamma^{ij} \gamma^{kl} - \gamma^{ik} g^{jl} \right )  n_l \right ]   \\
\end{eqnarray}
 The second term on the right hand side vanishes when contracting $n_l$ with the induced metric $\gamma^{kl}$ which is also included in the covariant derivative $\mathcal{D}_i$ defined on the boundary $^3 B$.  
 The remaining first term may or may not be useful, but it suffices to express the change of QLE as a flux through $B$. By means of Stokes' law the surface integral could be converted into a volume integral
of a quantity which superficially would look like the sought-after volume density of the field energy, but this is misleading and we presently see no advantage in doing so.
The term would contain covariant derivatives of the induced metric $\gamma_{ij}$ and therefore would still depend on the choice of $B$.   

In either case in the limit $r \longrightarrow 0$ only terms containing stress-energy are relevant in the presence of matter which as was pointed out before is a consequence of the absence of local energy-momentum.

Thus, the energy in eqn. \ref{E_self} is entirely due to stress-energy.

\subsection{Miscellaneous Values}

In the following diagrams the unreferenced QLE as given by eqn. \ref{Enoref} has been plotted for $m=1$ and various
values of $a$ and $Q$. Alternatively, one could consider the dimensionless ratio $E_1 / m$ which would depend on
the dimensionless quantities $a/m$ and $Q/e$ resulting in the same plots. 

\begin{figure}
\scalebox{0.4}{\includegraphics{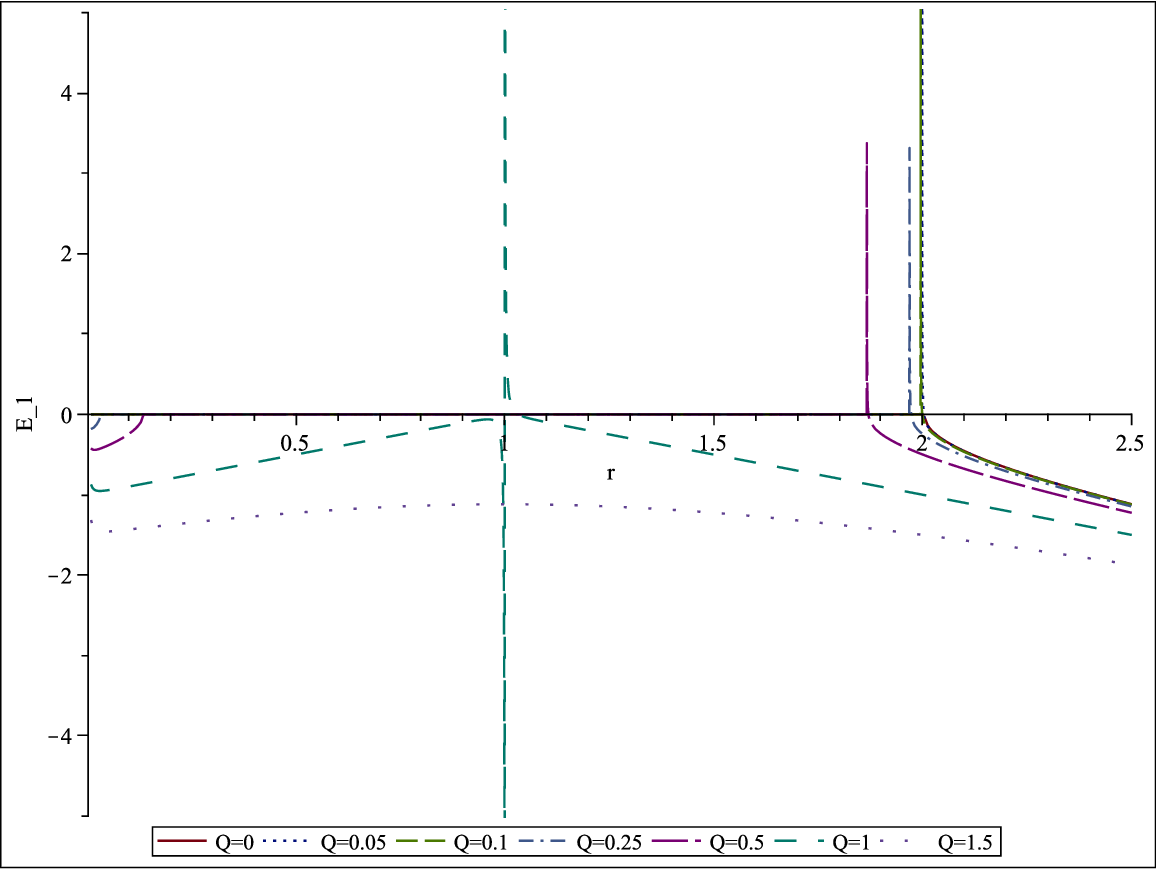}}
\caption{Real part of the unreferenced QLE given by eqn. \ref{Enoref} for $m=1$, $a=0.01$ and various values of $Q$}
\end{figure}

\begin{figure}
\scalebox{0.4}{\includegraphics{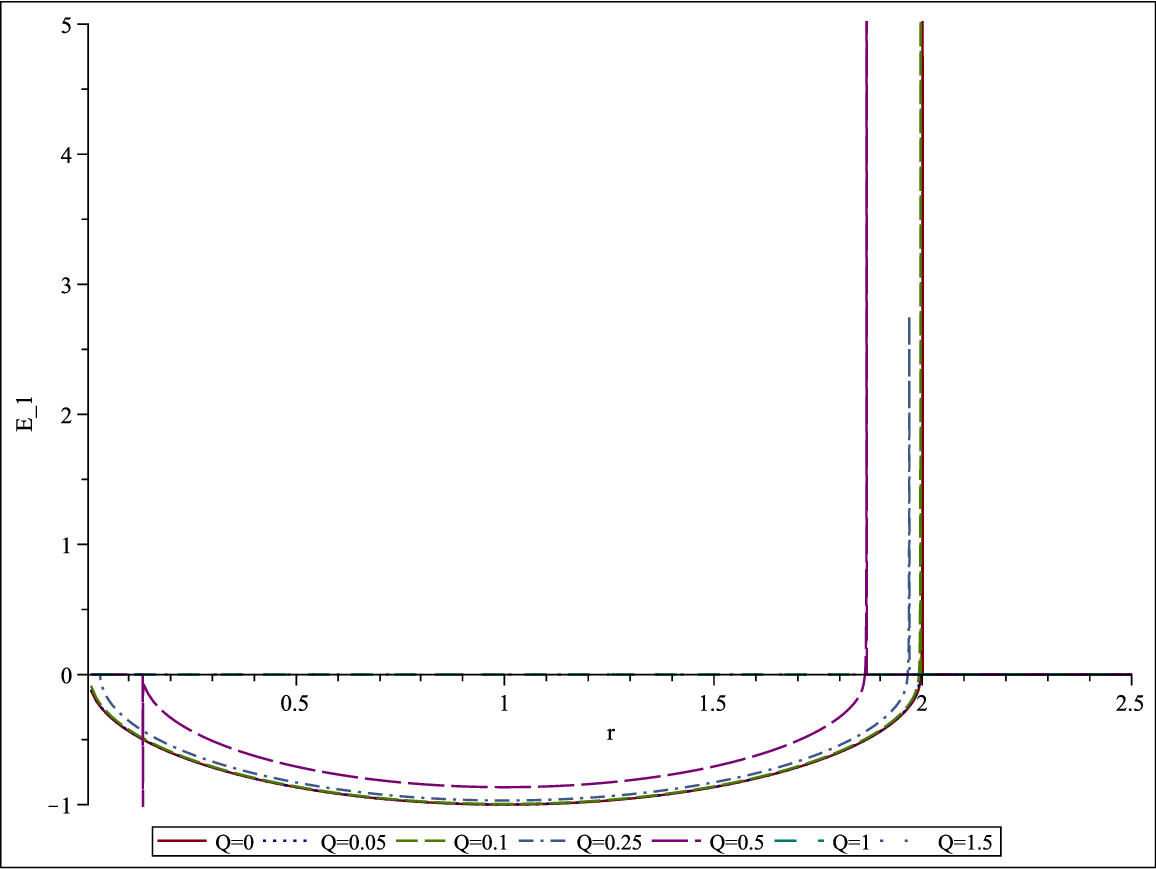}}
\caption{Imaginary part of the unreferenced QLE given by eqn. \ref{Enoref} for $m=1$, $a=0.01$ and various values of $Q$}
\end{figure}

\begin{figure}
\scalebox{0.4}{\includegraphics{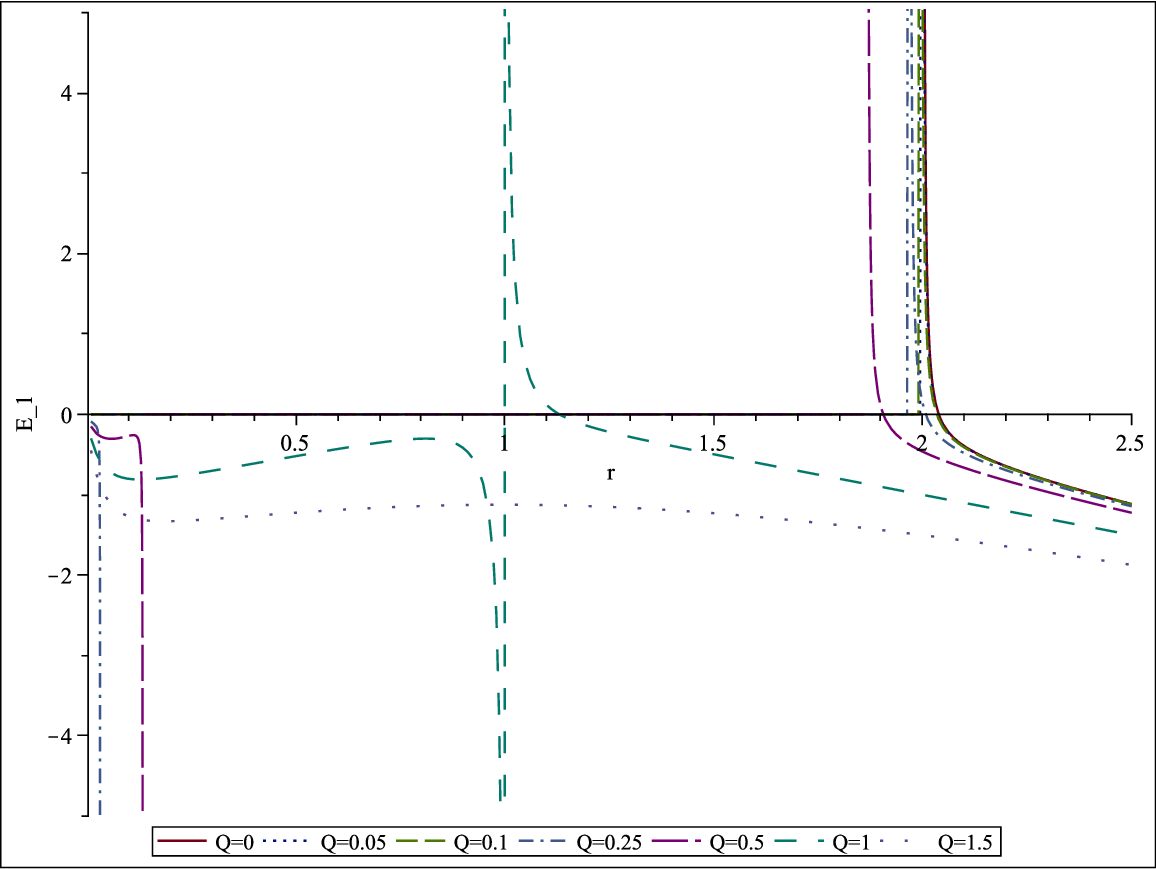}}
\caption{Real part of the unreferenced QLE given by eqn. \ref{Enoref} for $m=1$, $a=0.10$ and various values of $Q$}
\end{figure}

\begin{figure}
\scalebox{0.4}{\includegraphics{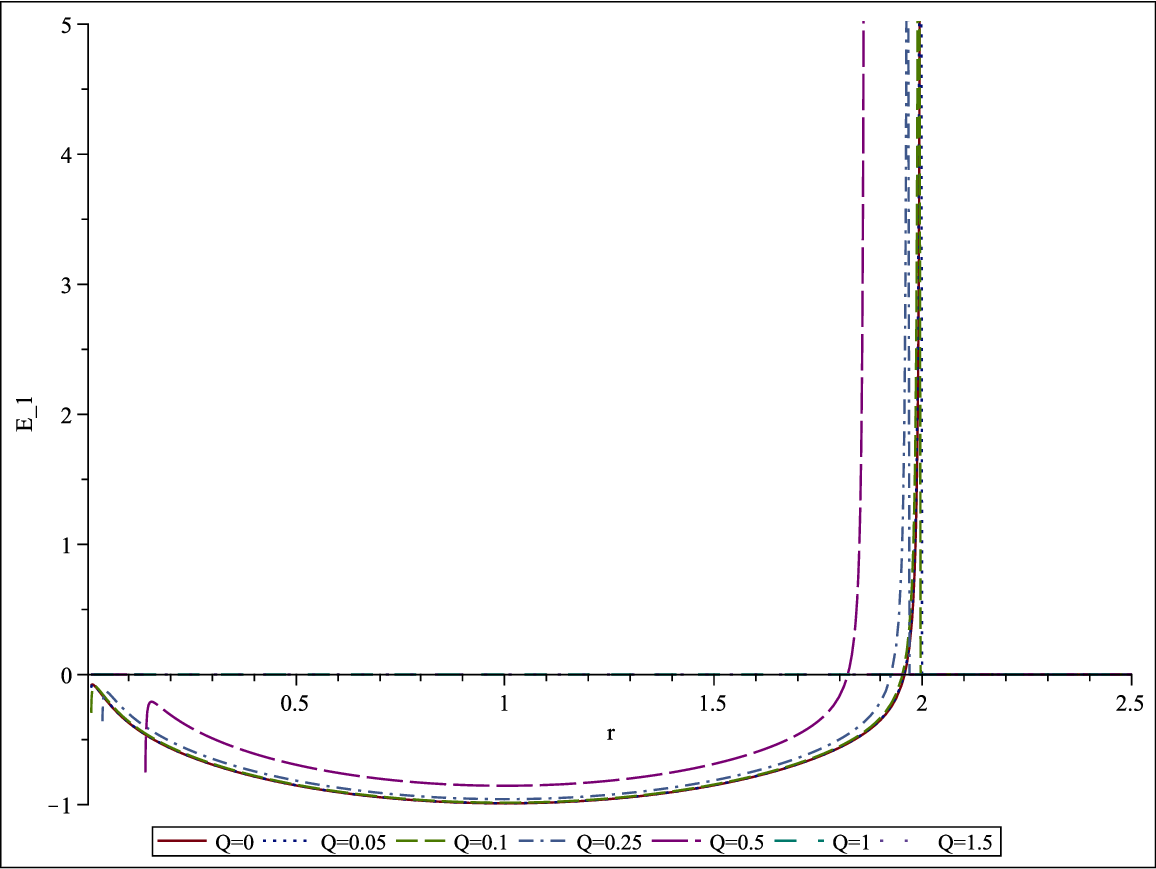}}
\caption{Imaginary part of the unreferenced QLE given by eqn. \ref{Enoref} for $m=1$, $a=0.10$ and various values of $Q$}
\end{figure}

\begin{figure}
\scalebox{0.4}{\includegraphics{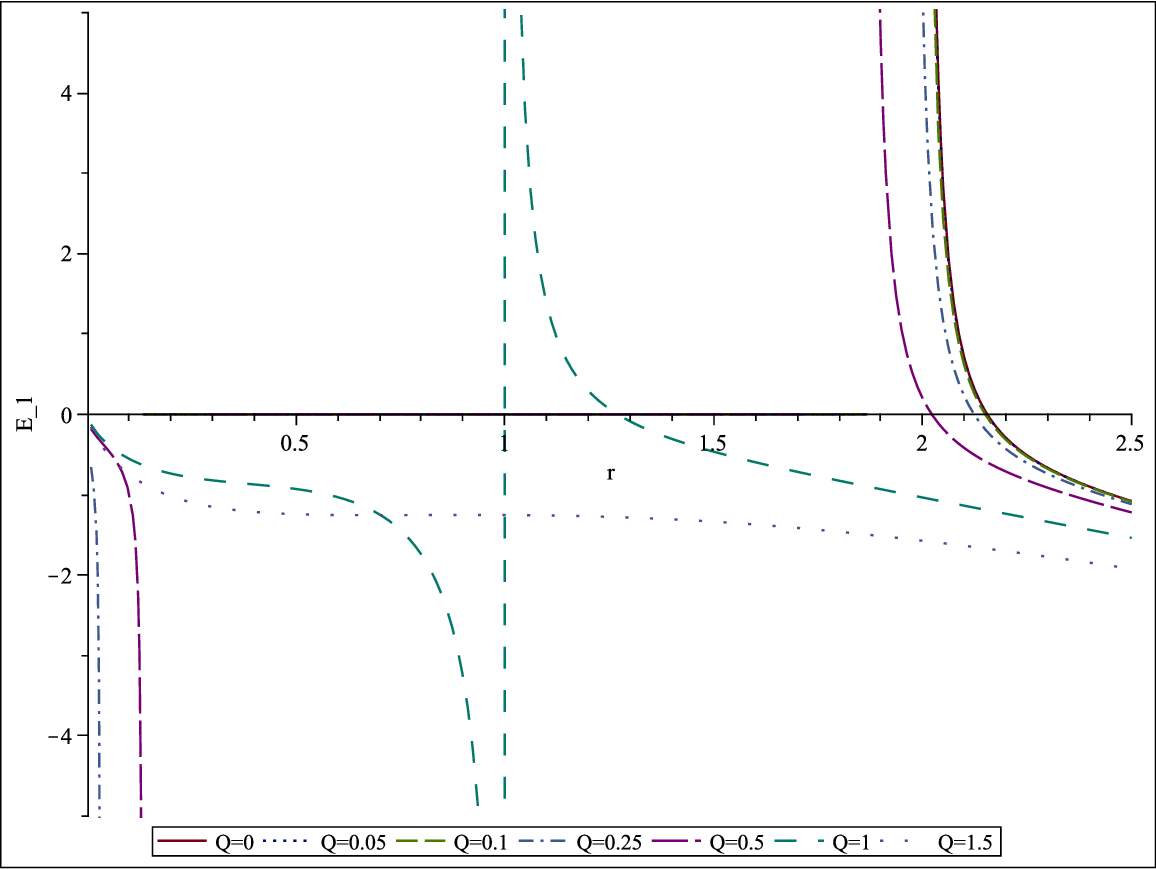}}
\caption{Real part of the unreferenced QLE given by eqn. \ref{Enoref} for $m=1$, $a=0.50$ and various values of $Q$}
\end{figure}

\begin{figure}
\scalebox{0.4}{\includegraphics{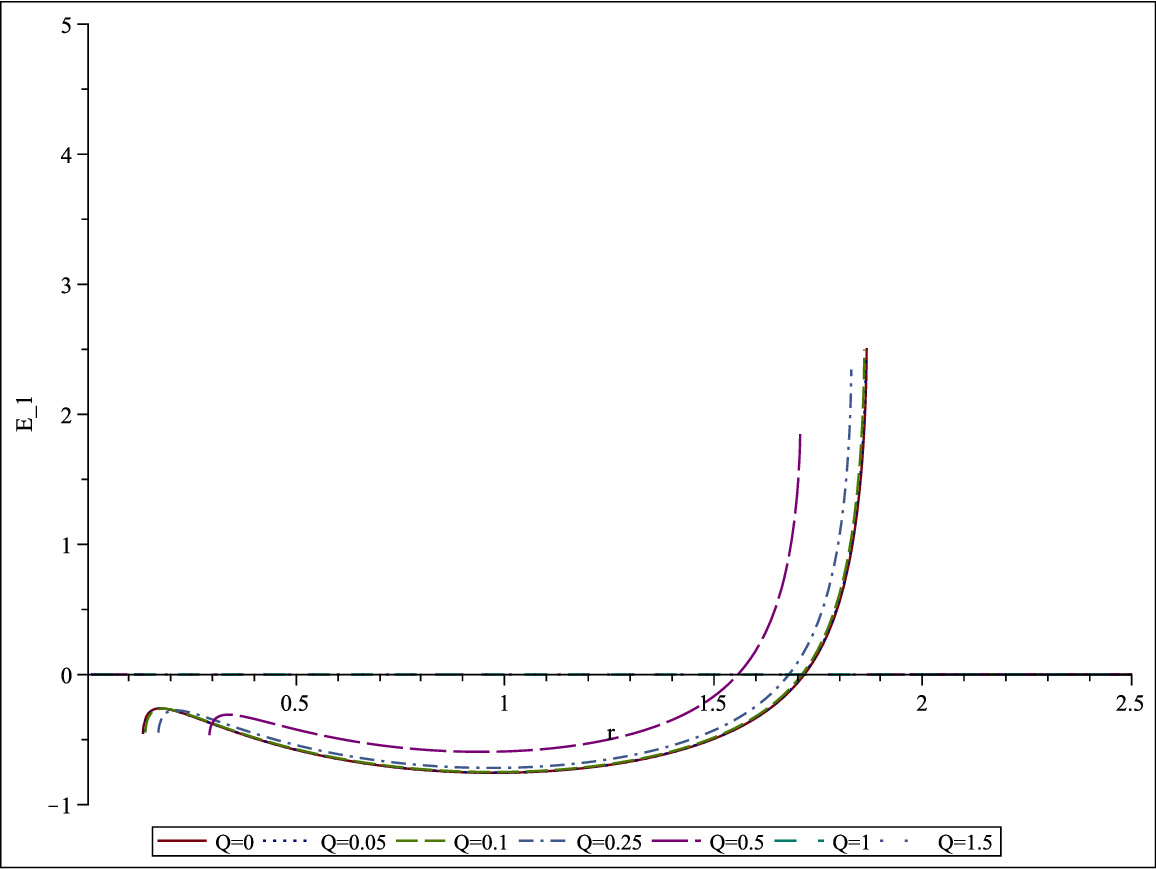}}
\caption{Imaginary part of the unreferenced QLE given by eqn. \ref{Enoref} for $m=1$, $a=0.50$ and various values of $Q$}
\end{figure}

\begin{figure}
\scalebox{0.4}{\includegraphics{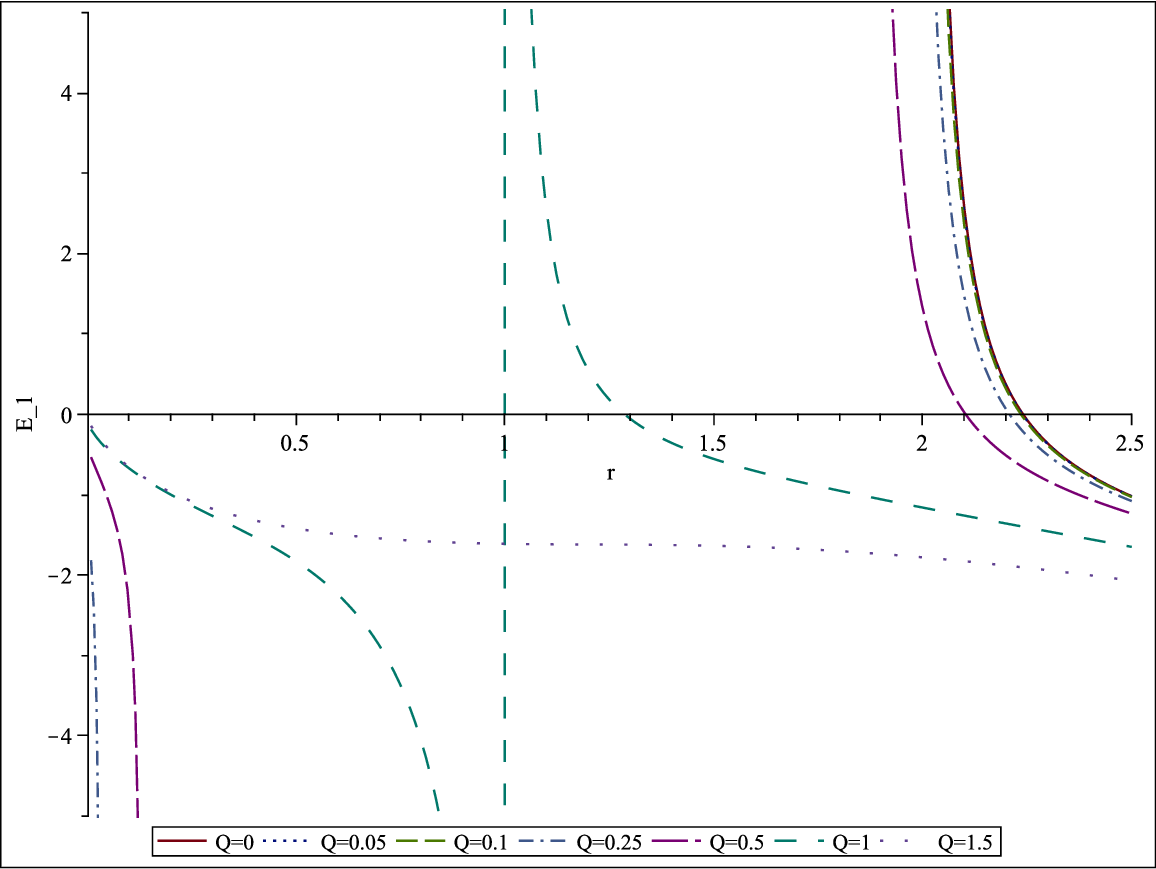}}
\caption{Unreferenced QLE given by eqn. \ref{Enoref} for $m=1$, $a=1.00$ and various values of $Q$. 
In this parameter range the QLE is entirely real.}
\end{figure}

\begin{figure}
\scalebox{0.4}{\includegraphics{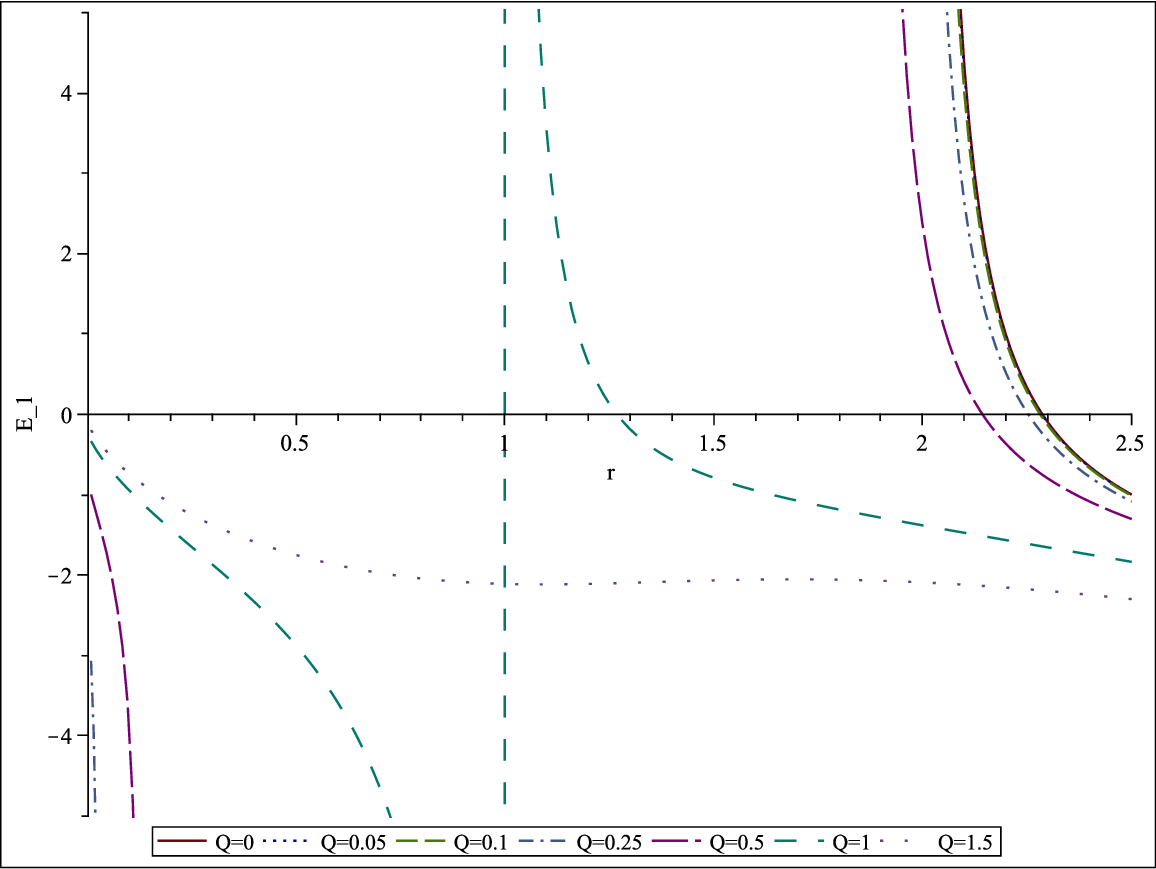}}
\caption{Unreferenced QLE given by eqn. \ref{Enoref} for $m=1$, $a=1.50$ and various values of $Q$. 
In this parameter range the QLE is entirely real.}
\end{figure}

In future work the presented computations should be repeated with a non-zero cosmological constant. 
An open problem is the presence of an electric quadrupole moment in the Kerr-Newman metric which may have no experimental basis. More general asymptotically flat spacetimes which
admit $g=2$ may solve this problem \cite{Kozameh_2008,PhysRevD.83.044023}. Suitable metrics for this purpose were found by Plebanski and Demianski \cite{PLEBANSKI197698,DEMIANSKI1972157,ExactSpacetimes}. 
Other obvious generalizations would include numbers of dimensions other than $3+1$. Current research indicates that
in ordinary general relativity the Kerr-Newman solution exists only in $3+1$ dimensions \cite{Adamo:2014baa,Bernard_blackholes,Myers:1986un} providing a possible argument
why stable matter cannot exist in a spacetime with a different number of dimensions.
Also, attempts should be made at recovering the exact value of the self-energy eqn. \ref{E_self}, e.g. by use of semi-classical quantization methods like path integral quantization along null geodesics \cite{York:2005kn}.

\acknowledgments
The author would like to acknowledge insightful discussion with James W. York, Jr. and Andrew P. Lundgren as well as a stimulating talk with subsequent discussion session given by Brandon Carter at Cornell University.

\bibliography{bib}
\bibliographystyle{h-physrev}

\appendix
\begin{widetext}

\section{Intermediate Results}
In the listing below $\chi = \cos \theta$ has been used to simplify the non-vanishing components of $\Theta_{\nu}^{\mu}$ and $\tau^{\mu \nu}$.

\begin{eqnarray*} 
\Theta_t^t = {\frac {{\chi}^{2}{a}^{4}m+{\chi}^{2}{a}^{2}m{r}^{2}+{Q}^{2}{a}^{2}r+{
Q}^{2}{r}^{3}-{a}^{2}m{r}^{2}-m{r}^{4}}{ \left( {a}^{2}{\chi}^{2}+{r}^
{2} \right) ^{2} \left( {Q}^{2}+{a}^{2}-2\,mr+{r}^{2} \right) }\sqrt {
{\frac {{Q}^{2}+{a}^{2}-2\,mr+{r}^{2}}{{a}^{2}{\chi}^{2}+{r}^{2}}}}} 
\end{eqnarray*} 

\begin{eqnarray*}
\Theta_\theta^\theta = -{\frac {r}{{a}^{2}{\chi}^{2}+{r}^{2}}\sqrt {{\frac {{Q}^{2}+{a}^{2}-2
\,mr+{r}^{2}}{{a}^{2}{\chi}^{2}+{r}^{2}}}}} 
\end{eqnarray*} 

\begingroup
\small
\begin{eqnarray*}
\Theta_\phi^\phi = -{\frac {-{\chi}^{4}{a}^{4}m+{\chi}^{4}{a}^{4}r+{\chi}^{2}{a}^{4}m-{
\chi}^{2}{a}^{2}m{r}^{2}+2\,{\chi}^{2}{a}^{2}{r}^{3}+{Q}^{2}{a}^{2}r+{
Q}^{2}{r}^{3}-{a}^{2}m{r}^{2}-2\,m{r}^{4}+{r}^{5}}{ \left( {a}^{2}{
\chi}^{2}+{r}^{2} \right) ^{2} \left( {Q}^{2}+{a}^{2}-2\,mr+{r}^{2}
 \right) }\sqrt {{\frac {{Q}^{2}+{a}^{2}-2\,mr+{r}^{2}}{{a}^{2}{\chi}^
{2}+{r}^{2}}}}} 
\end{eqnarray*} 
\endgroup

\begin{eqnarray*}
\Theta_t^\phi = {\frac { \left( {\chi}^{2}{Q}^{2}{a}^{2}r+{\chi}^{2}{a}^{4}m-{\chi}^{2
}{a}^{2}m{r}^{2}+{Q}^{2}{a}^{2}r+2\,{Q}^{2}{r}^{3}-{a}^{2}m{r}^{2}-3\,
m{r}^{4} \right) a \left( {\chi}^{2}-1 \right) }{ \left( {a}^{2}{\chi}
^{2}+{r}^{2} \right) ^{2} \left( {Q}^{2}+{a}^{2}-2\,mr+{r}^{2}
 \right) }\sqrt {{\frac {{Q}^{2}+{a}^{2}-2\,mr+{r}^{2}}{{a}^{2}{\chi}^
{2}+{r}^{2}}}}} 
\end{eqnarray*} 

\begin{eqnarray*}
\Theta_\phi^t = {\frac { \left( {\chi}^{2}{a}^{2}m+{Q}^{2}r-m{r}^{2} \right) a}{
 \left( {a}^{2}{\chi}^{2}+{r}^{2} \right) ^{2} \left( {Q}^{2}+{a}^{2}-
2\,mr+{r}^{2} \right) }\sqrt {{\frac {{Q}^{2}+{a}^{2}-2\,mr+{r}^{2}}{{
a}^{2}{\chi}^{2}+{r}^{2}}}}}
\end{eqnarray*} 

\begin{eqnarray*}
\tau^{tt} = {\frac {{\chi}^{2}{a}^{2}m-{\chi}^{2}{a}^{2}r-{a}^{2}m-{a}^{2}r-2\,{r}
^{3}}{\kappa\, \left( {a}^{2}{\chi}^{2}+{r}^{2} \right)  \left( {Q}^{2
}+{a}^{2}-2\,mr+{r}^{2} \right) }\sqrt {{\frac {{Q}^{2}+{a}^{2}-2\,mr+
{r}^{2}}{{a}^{2}{\chi}^{2}+{r}^{2}}}}} 
\end{eqnarray*} 

\begin{eqnarray*}
\tau^{\theta \theta} = -{\frac {m-r}{\kappa\, \left( {a}^{2}{\chi}^{2}+{r}^{2} \right) 
 \left( {Q}^{2}+{a}^{2}-2\,mr+{r}^{2} \right) }\sqrt {{\frac {{Q}^{2}+
{a}^{2}-2\,mr+{r}^{2}}{{a}^{2}{\chi}^{2}+{r}^{2}}}}} 
\end{eqnarray*} 

\begin{eqnarray*}
\tau^{\phi\phi} = {\frac {m-r}{\kappa\, \left( {\chi}^{2}-1 \right)  \left( {a}^{2}{\chi
}^{2}+{r}^{2} \right)  \left( {Q}^{2}+{a}^{2}-2\,mr+{r}^{2} \right) }
\sqrt {{\frac {{Q}^{2}+{a}^{2}-2\,mr+{r}^{2}}{{a}^{2}{\chi}^{2}+{r}^{2
}}}}} 
\end{eqnarray*} 

\begin{eqnarray*}
\tau^{t \phi} = \tau^{\phi t} = -{\frac {am}{\kappa\, \left( {a}^{2}{\chi}^{2}+{r}^{2} \right) 
 \left( {Q}^{2}+{a}^{2}-2\,mr+{r}^{2} \right) }\sqrt {{\frac {{Q}^{2}+
{a}^{2}-2\,mr+{r}^{2}}{{a}^{2}{\chi}^{2}+{r}^{2}}}}} 
\end{eqnarray*}

\section{Maple input code}
\begin{lstlisting}[language=C]
-I*(-r^3+3*m*r^2+(-2*m^2-Q^2-1/2*a^2)*r+m*(Q^2+1/2*a^2))*((Q^2+a^2-2*m*r+r^2)/(-2*m*r+Q^2+r^2))^(1/2)/(Q^2+a^2-2*m*r+r^2)^(1/2)/abs(a)*EllipticE(I*abs(a)/abs(r),(1/(-2*m*r+Q^2+r^2))^(1/2)*abs(r))-I*(-1/2*r^3-1/2*m*r^2+(2*m^2-1/2*a^2)*r-m*(Q^2+1/2*a^2))*((Q^2+a^2-2*m*r+r^2)/(-2*m*r+Q^2+r^2))^(1/2)/(Q^2+a^2-2*m*r+r^2)^(1/2)/abs(a)*EllipticF(I*abs(a/r),(1/(-2*m*r+Q^2+r^2))^(1/2)*abs(r))+2*(Pi*(a^2+r^2)^(1/2)*(-2*m*r+Q^2+r^2)*(piecewise(-1 < -(-Q^2+2*m*r-r^2)^(1/2)/abs(a),infinity*(signum(signum(Q^2+a^2-2*m*r+r^2)*(Q^2+a^2-2*m*r+r^2)^(1/2)*((-Q^2-a^2+2*m*r-r^2)/(Q^2-2*m*r))^(1/2)*(m-r)*(Q^2-2*m*r)/(-Q^2+2*m*r-r^2)^(3/4))-signum((Q^2+a^2-2*m*r+r^2)^(1/2)*((-Q^2-a^2+2*m*r-r^2)/(Q^2-2*m*r))^(1/2)*signum(Q^2+a^2-2*m*r+r^2)*(-1/(-Q^2+2*m*r-r^2)^(1/2))^(1/2)*(m-r)*(Q^2-2*m*r)/(-Q^2+2*m*r-r^2)^(1/2))),0)+piecewise((-Q^2+2*m*r-r^2)^(1/2)/abs(a) < 1,infinity*(signum(signum(Q^2+a^2-2*m*r+r^2)*(Q^2+a^2-2*m*r+r^2)^(1/2)*((-Q^2-a^2+2*m*r-r^2)/(Q^2-2*m*r))^(1/2)*(m-r)*(Q^2-2*m*r)/(-Q^2+2*m*r-r^2)^(3/4))-signum((Q^2+a^2-2*m*r+r^2)^(1/2)*((-Q^2-a^2+2*m*r-r^2)/(Q^2-2*m*r))^(1/2)*signum(Q^2+a^2-2*m*r+r^2)*(-1/(-Q^2+2*m*r-r^2)^(1/2))^(1/2)*(m-r)*(Q^2-2*m*r)/(-Q^2+2*m*r-r^2)^(1/2))),0))*(Q^2+a^2-2*m*r+r^2)^(1/2)-1/4*(r^2+abs(a)^2)*(m-r)*(Q^2-2*m*r+r^2+abs(a)^2))/(Q^2+a^2-2*m*r+r^2)^(1/2)/(a^2+r^2)^(1/2)/(-2*m*r+Q^2+r^2)
\end{lstlisting}

\end{widetext}

\end{document}